\documentclass[a4paper]{article}
\usepackage{graphicx}
\usepackage{graphics}
\usepackage{amssymb}
\usepackage{amsthm}
\usepackage{geometry}
\usepackage{txfonts}
\usepackage{mathdots}
\usepackage{natbib}

\begin{document}
\begin{center}
{\LARGE \textbf{Hotelling's test for highly correlated data}}\\
P. Bubeliny\\
\textsc{e-mail: bubeliny@karlin.mff.cuni.cz}\\
Charles University, Faculty  of  Mathematics  and
Physics, KPMS,  Sokolovska 83, Prague, Czech Republic, 18675.\\
\end{center}

\textbf{Abstract:} \textit{This paper is motivated by the analysis
of gene expression sets, especially by finding differentially
expressed gene sets between two phenotypes. Gene $\log_2$ expression
levels are highly correlated and, very likely, have approximately
normal distribution. Therefore, it seems reasonable to use
two-sample Hotelling's test for such data.  We discover some
unexpected properties of the test making it different from the
majority of tests previously used for such data. It appears that the
Hotelling's test does not always reach maximal power when all
marginal distributions are differentially expressed. For highly
correlated data its maximal power is attained when about a half of
marginal distributions are essentially different. For the case when
the correlation coefficient is greater than 0.5 this test is more
powerful if only one marginal distribution is shifted, comparing to
the case when all marginal distributions are equally shifted.
Moreover, when the correlation coefficient increases the power of
Hotelling's test increases as well.}

\section{Introduction}
\indent In many situations statisticians  need to test
multidimensional hypotheses. In a lot of cases components of
observed random vectors are highly dependent, which may change the
properties of the tests used. One of the examples of such data is
provided by gene expression levels. Gene expressions are highly
correlated between genes (see for example \citet*{kleb}). Moreover,
often the genes are investigated not just separately, but also as a
set of dependent genes. Therefore one has to deal with
multidimensional hypotheses and in order to test such hypotheses,
gene sets should be expressed differentially. The most popular tests
for gene sets are Hotelling's test, N-test and tests derived from
marginal $t$-statistics. In the papers \citet{ackerman},
\citet{glazko}, an approach to comparing these test in various
situations was made. Our goal is not to make another comparison, but
rather to describe some interesting properties of the Hotelling's
test which seems to be unexpected.

\section{Hotelling's test}

\indent One of the most well known tests is $t$-test. Hotelling's
test is an multidimensional extension of $t$-test. Similar to
$t$-test, we can consider both one-sample and two-sample Hotelling's
test. One-sample case deals with the hypothesis that the expected
value of a sample from multidimensional normal distribution is equal
to some given vector. In the two-sample case it deals with the
hypothesis of the equality of expected values of two samples from
multidimensional normal distributions (with the equal covariance
structure). In this paper we will focus on the two-sample
Hotelling's test.\\
\indent Suppose we have two independent samples (of sizes $n_x$ and
$n_y$, respectively) from two $n$-dimensional normal distributions
with identical covariance matrices equal to $\Sigma$. In other
words, we consider $X_1,...,X_{n_{x}}$ as i.i.d random vectors
having $N_n(\mu_{x},\Sigma)$ and $Y_1,...,Y_{n_{y}}$ as i.i.d random
vectors having $N_n(\mu_{y},\Sigma)$ ($X_i$ and $Y_j$ are
independent for all $i=1,...,n_x;j=1,...,n_y$). For simplicity we
assume that $n<n_x+n_y-1$. Our goal is to test the hypothesis $H:$
$\mu_x=\mu_y$ against alternative $A:$ $\mu_x\neq\mu_y$. For this we
use Hotelling's test based on the statistic
\begin{equation}\label{hots}
T^2=\frac{n_x
n_y}{n_x+n_y}(\bar{X}-\bar{Y})^TS^{-1}(\bar{X}-\bar{Y}),
\end{equation}
where $\bar{X}=\frac{1}{n_x}\sum_{i=1}^{n_x} X_i$;
$\bar{Y}=\frac{1}{n_y}\sum_{i=1}^{n_y} Y_i$ and
$S=\frac{\sum_{i=1}^{n_x}(X_i-\bar{X})(X_i-\bar{X})^T+\sum_{i=1}^{n_y}
(Y_i-\bar{Y})(Y_i-\bar{Y})^T}{n_x+n_y-2}$. $T^2$ is related to the
$F$-distribution by
\begin{equation}\label{fapr}
\frac{n_x+n_y-n-1}{n(n_x+n_y-2)}T^2\thicksim F(n,n_x+n_y-n-1).
\end{equation}
For more details about Hotelling's test see, for example,
\citet{chat}. We made the assumption $n<n_x+n_y-1$ for two reasons.
For $n\geq n_x+n_y-1$ the estimate $S$ of $\Sigma$ results in an
irregular matrix, so that $S^{-1}$ does not exist and moreover
numerator of (\ref{fapr}) is non-positive as well as the degree of
freedom of the $F$-distribution. In such situations it is possible
to use some pseudo-inversion of $S$ and in order to estimate
$p$-value of $H$, we can use permutations of
$(X_1,...,X_{n_x},Y_1,...,Y_{n_y})$.

\section{Hotelling's test for strongly dependent data}

\indent As it was mentioned above, genes are highly dependent and we
will suppose that their log$_2$ expression levels have approximately
normal distributions. Many papers work with gene sets (for example
\citet{barry}) instead of genes alone and therefore deal with
multidimensional hypotheses. It seems to be reasonable to use Hotelling's test in this situation.\\
\indent Assume that we have two multidimensional samples and need to
test the hypothesis suggesting the equality of expected values in
these two samples. Assume for simplicity that all elements on the
main diagonal of the covariance matrix $\Sigma$ for both samples are
equal to 1 and all other elements are equal to $\rho>0$, i.e.
\[ \Sigma =\left( \begin{array}{ccccc}
1 & \rho &\rho& ...&\rho \\
\rho & 1 & \rho&... &\rho\\
...&...&...&...&... \\
\rho & ...&... & \rho &1\end{array} \right).\]

\begin{figure}[h!]
\begin{center}
\includegraphics[width=12cm]{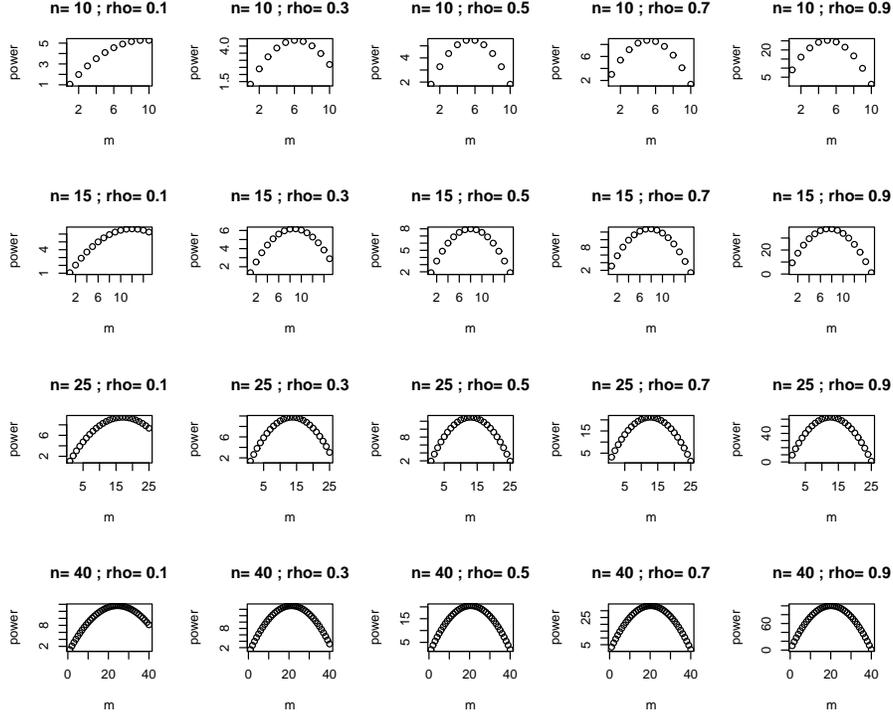}
\end{center}
\caption{Plots of $T^{*2}$ for $n=10$, 15, 25, 40; $\rho=0.1$, 0.3,
0.5, 0.7, 0.9; and $m=1,...,n$. Notice: each plot is differently
scaled!} \label{ruzm}
\end{figure}

Further on, we assume that $\mu_x=(0,...,0)^T$, but $\mu_y$ has
first $m$ elements equal to 1 and the others equal to 0, i.e.
\[\mu_y=\big(\underbrace{1,...,1}_m,\underbrace{0,...,0}_{n-m}\big)^T.\]
For large $n_x$ and $n_y$ the matrix $\Sigma$ and its estimate $S$
are approximately the same as well as the differences between the
expected values ($\mu_x-\mu_y$) and between the mean values
($\bar{X}-\bar{Y}$). When dialing with real data, $n_x$ and $n_y$
might not be large enough, but for theoretical reasons we may use
the approximations $S\approx\Sigma$ and
$\bar{X}-\bar{Y}\approx\mu_x-\mu_y$. In this case
$S^{-1}\approx\Sigma^{-1}$, that is
\[S^{-1}\approx \Sigma^{-1} =\left( \begin{array}{ccccc}
\alpha & -\beta &-\beta& ...&-\beta \\
-\beta & \alpha & -\beta&... &-\beta\\
...&...&...&...&... \\
-\beta & ...&... & -\beta &\alpha\end{array} \right),\] where
$\alpha=\frac{(1+(n-2)\rho)}{(1-\rho)(1+(n-1)\rho)}$ and
$\beta=\frac{\rho}{(1-\rho)(1+(n-1)\rho)}$. For fixed $n_x$ and
$n_y$ we can consider the fraction $\frac{n_xn_y}{n_x+n_y}=k$ of
Hotelling's statistic (\ref{hots}) as a normalizing constant. Let us
denote $T^{*2}$ Hotelling's statistic with $\Sigma^{-1}$ instead of
$S^{-1}$ and $\mu_x-\mu_y$ instead of $\bar{X}-\bar{Y}$ divided by
the constant $k$. Therefore, we have
\[ T^2/k\approx T^{*2}=(\mu_x-\mu_y)^T\Sigma^{-1}(\mu_x-\mu_y) \]
\[=\big(\underbrace{1,...,1}_m,\underbrace{0,...,0}_{n-m}\big)\left(
\begin{array}{ccccc}
\alpha & -\beta &-\beta& ...&-\beta \\
-\beta & \alpha & -\beta&... &-\beta\\
...&...&...&...&... \\
-\beta & ...&... & -\beta &\alpha\end{array} \right) \left(
\begin{array}{c}
1 \\
...\\
1 \\
0\\
...\\
0\\\end{array} \right)\]
\begin{equation}\label{sim}
=m\alpha-(m^2-m)\beta=\frac{m(1+(n-2)\rho)-m(m-1)\rho}{(1-\rho)(1+(n-1)\rho)}
=\frac{m(1+(n-m-1)\rho)}{(1-\rho)(1+(n-1)\rho)}.
\end{equation}
Let us note that it does not matter if $\mu_y$ consists of ones and
zeros or equals to a constant $a$ and zeros. In the latter case,
statistic $T^{*2}$ would be multiplied by $a^2$. Now we will work
with  statistic $T^{*2}$ and investigate its behavior.\\
\indent If we changed $m$ to $m+1$ (meaning that we add one more
different marginal distribution) we would expect that the statistic
$T^{*2}$ increases and that so does the power of Hotelling's test.
We need to check if it is indeed the case. For better understanding
let the number of ones in $\mu_y$ be the index of $T^{*2}$ (we will
write it only when it is needed). Now we change $m$ to $m+1=h$ and
we have
\[
T^{*2}_{m+1}=T^{*2}_{m}+\alpha-2m\beta.
\]
If we expected that $T^{*2}$ is an increasing function of $m$ then
$\alpha-m^2\beta$ should be greater then zero. But we have
\[
\alpha-2m\beta=\frac{1+(n-2)\rho}{(1-\rho)(1+(n-1)\rho)}-
\frac{2m\rho}{(1-\rho)(1+(n-1)\rho)}=\frac{1+(n-2m-2)\rho}{(1-\rho)(1+(n-1)\rho)}.
\]
Since the denominator is greater than zero, then $\alpha-2m\beta>0$
only if $\frac{1}{2m+2-n}=\frac{1}{2h-n}>\rho$. It means that for
not very small values of $\rho$'s and $m>\frac{n}{2}-1$ the
statistic $T^{*2}$ is a decreasing function of $m$. This means that
maximal power of Hotelling's test (as a function of $m$) is not
always attained for $m=n$ but for $\rho$'s which are not very small
we have maximal power for $m$ near $\frac{n}{2}$. Some examples of
the behavior of $T^{*2}$ as
a function of $m$ are illustrated on figure \ref{ruzm}.\\
However, this issue is not the only one that is surprising about
Hotelling's test. Now we look if $T^{*2}_1$ is always lower than
$T^{*2}_n$. It is the case when one different marginal distribution
influences more than all $n$ different distributions. So we need to
compare $\alpha$ with $n\alpha-n(n-1)\beta$. We have
\[
T^{*2}_1-T^{*2}_n=\alpha-n\alpha+n(n-1)\beta=(n-1)\frac{(1-2\rho)}{(1-\rho)(1+(n-1)\rho)}.\]
So $T^{*2}_1-T^{*2}_n<0$ only if $\rho<0.5$. Therefore we can say
that for $\rho>0.5$ Hotelling's test has better power for
alternative with only one marginal shift than for alternative that
all marginal distributions are equally shifted. It can be seen from
figure \ref{ruzm} as well. Moreover, the statistic $T^{*2}$ is an
increasing function of $\rho$, that may seem surprising as well.

\section{Hotelling's test for two-dimensional data}

\begin{figure}[h!]
\begin{center}
\includegraphics[width=12cm]{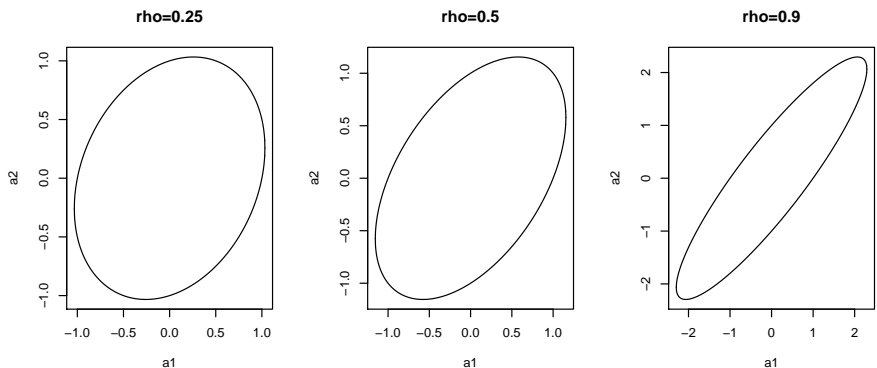}
\end{center}
\caption{Plots of solutions of equation (\ref{obr}) for
two-dimensional case for $rho=0.25$;0.5;0.9. Notice: each plot is
differently scaled!}

\label{elip2d}
\end{figure}

\indent Let us look at Hotelling's test in the two-dimensional case.
As in the previous case, we will consider the two-sample problem,
but now we will generalize the difference of expected values of
these two samples. Suppose that $\mu_x-\mu_y=(a_1,a_2)$ and that the
covariance matrix is
\[ \Sigma =\left( \begin{array}{cc}
1 &\rho \\
\rho &1\end{array} \right).\] Then inverse of $\Sigma$ is the matrix
with diagonal elements $\alpha=\frac{1}{(1-\rho)(1+\rho)}$ and
off-diagonal elements $-\beta=\frac{-\rho}{(1-\rho)(1+\rho)}$. Then
\[
T^{*2}=\alpha a_1^2+\alpha a_2^2 - 2\beta a_1a_2.
\]
First we consider that $a_1=1$ and $a_2=0$. Then $T^{*2}=\alpha$.
Now we will investigate for which $a_1,a_2\in R$ statistic
$T^{*2}=\alpha$. That is, we need to solve an equation
\begin{equation}\label{2dim}
\alpha a_1^2+\alpha a_2^2 - 2\beta a_1a_2=\alpha.
\end{equation}
After dividing both sides of equation (\ref{2dim}) by $\alpha$ we
get
\begin{equation}\label{root}
a_1^2 + a_2^2- 2\rho a_1a_2-1=0.
\end{equation}
For fixed $a_1$ equation (\ref{root}) is quadratic in $a_2$ with the
roots
\[
a_{2_{1,2}}=\frac{2\rho a_1\pm\sqrt{(2\rho a_1)^2-4(a_1^2-1)}}{2}.
\]
It is defined only if $(2\rho a_1)^2-4(a_1^2-1)\geq0$, i.e. for
$|a_1|\leq\sqrt{\frac{1}{1-\rho^2}}$. Some plots of the solutions of
the equation (\ref{root}) for different values of the correlation
coefficient $\rho$ are given on figure \ref{elip2d}. We can see that
the plots of these solutions produce elliptic curves. Let us rotate
these ellipses by the angle $\varphi=\Pi/4$ clockwise. To do this,
we use transformation
\[
a_1=x\cos\varphi-y\sin\varphi=\frac{\sqrt{2}}{2}x-\frac{\sqrt{2}}{2}y,
\]
\[
a_2=x\sin\varphi+y\cos\varphi=\frac{\sqrt{2}}{2}x+\frac{\sqrt{2}}{2}y,
\]
where $x$ and $y$ are new rotated coordinates. After substitution
into (\ref{root}) it gives
\[
(\frac{\sqrt{2}}{2}x-\frac{\sqrt{2}}{2}y)^2+(\frac{\sqrt{2}}{2}x+\frac{\sqrt{2}}{2}y)^2
-2\rho(\frac{\sqrt{2}}{2}x-\frac{\sqrt{2}}{2}y)(\frac{\sqrt{2}}{2}x+\frac{\sqrt{2}}{2}y)
\]
\[
=x^2(1-\rho)+y^2(1+\rho)=\frac{x^2}{a^2}+\frac{y^2}{b^2}=1,
\]
where $a=\sqrt{\frac{1}{1-\rho}}$ and $b=\sqrt{\frac{1}{1+\rho}}$
are respectively the major radius and the minor radius of the
ellipse. Since $a>b$, the Hotelling's test has the weakest power in
the direction of $a_1=a_2$, while the fastest increase of its power
is observed towards the direction of  $a_1=-a_2$. For example, for
$\rho=0.9$ we have $a=3.162$ and $b=0.725$. It means that for
$a_1=a_2=\sqrt{\frac{3.162^2}{2}}=2.236$ Hotelling's test has
approximately the same power as for $a_1=1$, $a_2=0$ (or for
$a_1=-a_2=\sqrt{\frac{0.725^2}{2}}=0.513$ as well). So, if there is
only one marginal distribution shifted by one unit, then the power
of Hotelling's test is approximately the same as if both marginal
distribution were equally shifted (in the same direction) by 2.236
units (for the shift in opposite direction it should be only 0.513
unit). These results are in contradiction with other
multidimensional tests. For example, consider the test based on
marginal $t$-statistics. The power of this test is higher if both
distributions are shifted by the same amount (both $t$-statistics
are "large", not depending on direction of shift) than if there was
only one marginal distribution shifted (one $t$-statistic is "near"
zero).

\section{Theory and reality}

\begin{figure}[h!]
\begin{center}
\includegraphics[width=12cm]{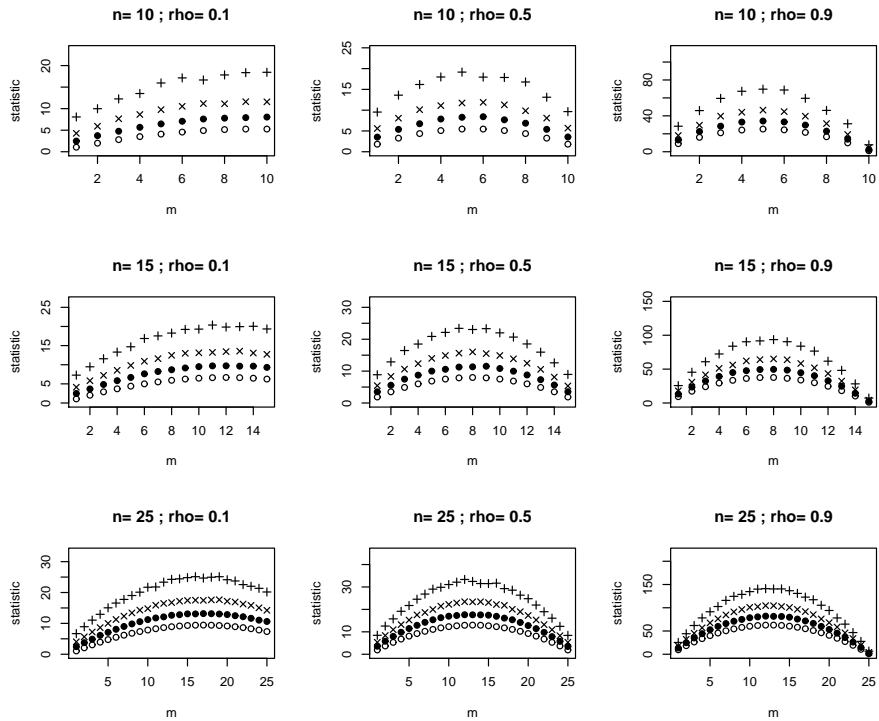}
\end{center}
\caption{Comparisons of theoretical statistics $T^{*2}$ and real
Hotelling's statistic $T^2/k$ for number of genes $n=10$ $15$, $25$
(from the top to the bottom); for correlation coefficient
$\rho=0.1$, $0.5$, $0.9$ (from the left to the right) and number of
observations in each sample $n_x=n_y=n$ (denoted by '+'),
$n_x=n_y=1.4n$ (denoted by 'x') and $n_x=n_y=2.4n$ (denoted by
'$\bullet$'). The theoretical statistic $T^{*2}$ is denoted by
'$\circ$'. Number of different marginal distribution $m$ is set from
one to $n$. Notice: each plot is differently scaled!} \label{difH}
\end{figure}

\indent The analytical results obtained above should be verified by
checking if actual Hotelling's test outcomes correspond to the
analytical results regarding real data. In this section we will
compare the behavior of theoretical Hotelling's statistic $T^{*2}$
with real Hotelling's statistic $T^2$. For large $n_x$ and $n_y$ we
assumed that $T^{*2}\approx T^2/k$, where $k=\frac{n_x
n_y}{n_x+n_y}$. Constant $k$ changes as $n_x$ and $n_y$ change. It
is reasonable to divide Hotelling's statistic $T^2$ by $k$ instead
of multiplying $T^{*2}$ by $k$ in order to be able to compare how do
$T^2$ and $T^{*2}$ differ for various $n_x$ and $n_y$.\\ \indent In
order to compare the actual results with the analytical ones, we did
the following simulations. All data were simulated from
$n$-dimensional normal distributions. Consider three different
values for the number of genes in a gene set. We take $n=10$, $n=15$
and $n=25$. All simulations were performed for three different
values of the correlation coefficient $\rho$\,:\,\, $\rho=0.1$,
$\rho=0.5$ and $\rho=0.9$. In order to compare the behavior of
Hotelling's test for various sizes of samples we took three choices
of $n_x$ and $n_y$:\,\, $n_x=n_y=n$,\,$n_x=n_y=1.4n$ and
$n_x=n_y=2.4n$. The value $m$ which is the number of false marginal
distributions varies from one to $n$. The shift value for each of
the different marginal distributions is set to one. The theoretical
Hotelling's statistic is calculated according to (\ref{sim}). Real
Hotelling's statistic is estimated from $1000$ simulations for each
case (as the mean of $T^2/k$ obtained from the simulations).\\
\indent Plots of our simulated cases are shown on figure \ref{difH}.
We can see that for all simulated situations, the shapes of real and
theoretical Hotelling's statistics are similar. The only difference
is in the heights of these curves. For small $n_x$ and $n_y$
statistic $T^2$ has higher values than for large $n_x$ and $n_y$.
The reason for that stems from the inaccurate estimates of the
expected values and of the covariance matrix. However, we observe
that with the increase of $n_x$ and $n_y$, statistic $T^2/k$ goes to
$T^{*2}$ relatively fast. Therefore, the behavior of Hotelling's
test for real data is expected to be very similar to the behavior of
statistic $T^{*2}$.\\
\indent In previous section we saw that for the two-dimensional case
the plotted shifts with equal values of the power of theoretical
Hotelling's test form elliptic curves. Hotelling's statistics $T^2$
are random variables. Therefore, we can only estimate if their
expected values form elliptic curves when plotted. To check this we
did following simulations. Instead of calculating the shifts for
which Hotelling's test has equal powers, we took the points provided
by the elliptic curves observed for theoretical Hotelling's
statistics. For each pair of these points $(a_1,a_2)$ we did 1000
simulations and calculated Hotelling's statistic. We estimated the
expected value E$T^2/k$ as the mean for these 1000 repetitions. We
divided Hotelling's statistics by $k$ for better understanding how
fast these statistics go to $T^{*2}$. We did this simulation for the
values of the correlation coefficient $\rho=0.3$ and $\rho=0.9$ and
as the number of observations in each sample we took $n_x=n_y=5$,
$n_x=n_y=10$ and $n_x=n_y=20$. Results of our simulation are given
in Table 1. We observe that estimated mean values of
$\overline{T^2/k}$ are not very different, that they go to $T^{*2}$
and that their variance decreases with increasing number of
observations. Clearly, these points form elliptic curves. Hence, we
can claim that the real Hotelling's test behaves very similar to the
theoretical one and the theory derived for the theoretical test
holds for the real Hotelling's test as well.
\begin{table}[ht]
\begin{center}
\caption{Results of simulations of two-dimensional adjusted
Hotelling's statistics $T^2/k$ with $n_s=n_x=n_y$ observations for
each sample and correlation coefficient $\rho$. $T^{*2}$ stands for
theoretical Hotelling's statistics and $(a_1,a_2)$ is difference
between expected values $\mu_x-\mu_y$ of these samples. On bottom
line is the estimate of variance of each column.} \label{tab}
\begin{tabular}{|c|c|ccc||c|c|ccc|}
  \hline
\multicolumn{2}{|c|}{$T^{*2}=1.0989$}&\multicolumn{3}{c||}{$\rho=0.3$}&
\multicolumn{2}{|c|}{$T^{*2}=5.2632$}&\multicolumn{3}{c|}{$\rho=0.9$}\\
\hline
$a_1$&$a_2$&$n_s=5$&$n_s=10$&$n_s=20$&$a_1$&$a_2$&$n_s=5$&$n_s=10$&$n_s=20$\\
  \hline
-0.84 & 0.35 & 3.12 & 1.74 & 1.35 & -1.83 & -1.05 & 9.58 & 6.72 & 5.96 \\
-0.63 & 0.61 & 3.03 & 1.81 & 1.42 & -1.38 & -0.44 & 9.55 & 6.51 & 5.96 \\
-0.42 & 0.79 & 3.04 & 1.82 & 1.39 & -0.92 & 0.09 & 9.55 & 6.65 & 5.99 \\
-0.21 & 0.92 & 3.00 & 1.75 & 1.42 & -0.46 & 0.57 & 9.62 & 6.93 & 5.98 \\
0.00 & 1.00 & 3.03 & 1.72 & 1.42 & 0.00 & 1.00 & 9.10 & 6.99 & 5.83 \\
0.21 & 1.04 & 3.04 & 1.74 & 1.36 & 0.46 & 1.39 & 9.74 & 6.78 & 5.99 \\
0.42 & 1.04 & 3.01 & 1.87 & 1.39 & 0.92 & 1.74 & 10.11 & 6.75 & 5.86 \\
0.63 & 0.99 & 3.00 & 1.79 & 1.40 & 1.38 & 2.04 & 9.36 & 6.87 & 5.85 \\
0.84 & 0.85 & 3.32 & 1.81 & 1.41 & 1.83 & 2.25 & 10.21 & 6.87 & 5.96 \\
1.05 & 0.35 & 3.35 & 1.85 & 1.36 & 2.29 & 2.09 & 9.94 & 6.85 & 5.97 \\
\hline \multicolumn{2}{|c}{var:}& 0.0176& 0.0025& 0.0007&
\multicolumn{2}{|c}{var:}& 0.1133& 0.0202& 0.0039\\
\hline
\end{tabular}
\end{center}
\end{table}
\section{Discussion}
\indent In this paper we have discovered that two-sample Hotelling's
test (for testing the equality of the expected values of two samples
from multidimensional normal distribution with equal covariance
structure) has some unexpected properties. At first sight, one could
expect that with a larger number of false marginal distributions the
power of this test increases. But we have discovered that this is
not true in general. For highly correlated and high dimensional data
(such as data sets of gene expressions) maximal power of Hotelling's
test is reached when
 only about one half of the marginal distributions are shifted. We
have found out that when the correlation inside the sample is
greater than 0.5, then the Hotelling's test can have a better power
if only one marginal distribution is different, as opposed to the
case when all marginal hypotheses are false. Moreover, the power of
Hotelling's test increases for higher correlations. That observation
may seem somewhat unexpected as well. We have investigated
Hotelling's test in detail in two-dimensional case. We have found
that properties of this test are much different from ones of the
tests based on marginal $t$-statistic. All reasonable tests based on
marginal $t$-statistic do not depend on the direction of the shift.
But the power of Hotelling's test increases very slowly if both of
the marginal distributions are equally shifted and increases much
faster if marginal distributions are shifted in opposite directions.
Moreover, alternatives with equal values of the power form
ellipsoids.

\section*{Acknowledgments}
\indent The author thanks Prof. Lev Klebanov, DrSc. for valuable
comments, remarks and overall help. The work was supported by the
grant SVV 261315/2010.

\bibliographystyle{elsarticle-harv}

\end{document}